\documentclass[conference]{IEEEtran}

\usepackage{amssymb}
\usepackage{array}
\usepackage{multirow}
\usepackage{graphicx}	
\usepackage{amsfonts} 
\usepackage[ruled,vlined,linesnumbered]{algorithm2e}

\begin{document}
\title{ Revocation Management in Vehicular Ad-hoc Networks}

\author{\IEEEauthorblockN{Francisco Mart\'in-Fern\'andez, Pino Caballero-Gil and C\'andido Caballero-Gil}
\IEEEauthorblockA{Department of Computer Engineering\\
University of La Laguna\\
38271 La Laguna Tenerife, Spain\\
Email: \{fmartinf, pcaballe, ccabgil\}@ull.edu.es}
}

\maketitle

\begin{abstract}
This paper describes a solution for the efficient management of revocation in vehicular ad-hoc networks, for both certificate-based and identity-based authentication. It proposes the use of an authenticated data structure based on a dynamic hash tree, which is a perfect $k$-ary tree, together with a new version of the SHA-3 hash function. This combination allows optimizing search and insertion operations in the tree. Consequently, the proposal is very useful both when vehicular networks are widely used, and in urban environments. Simulation results are promising and confirm this hypothesis.
\end{abstract}

\IEEEpeerreviewmaketitle

\section{Introduction}
Authentication is a crucial requirement for any communication network. On the one hand, an efficient way  to authenticate legitimate and honest  nodes is necessary. On the other hand,  being able to exclude compromised nodes  is fundamental to guarantee trustworthiness of network services. 

When communication security is based on  public-key  cryptography, a central problem is to guarantee that a particular public key is  authentic and valid. The traditional approach to this problem is through public-key certificates emitted by  a Public-Key Infrastructure (PKI), in which a Certificate Authority (CA) certifies ownership and validity of public-key certificates.  This solution presents many   difficulties because 
the issues associated with certificate management are quite complicated and expensive. A different approach  is the so-called Identity-Based Cryptography (IBC), where each user's public key is his/her public IDentity (ID) so that  the need for public-key certificates is eliminated. 

In order to use any public-key cryptosystem  in practice, an efficient revocation mechanism is  necessary because private keys may become compromised. Traditionally, this problem has been solved through a centralized approach  based on the existence of a Trusted Third Party (TTP), which  is usually a CA  distributing  the so-called Certificate Revocation Lists (CRLs) that can be seen as blacklists of revoked certificates. Alternatively, some authors have proposed an  approach based on hash trees as  Authenticated Data Structures (ADSs) for  a more efficient management of  certificate revocation.

Vehicular Ad-hoc NETworks (VANETs) are self-organizing networks built up from moving vehicles that communicate with each other mainly to prevent adverse circumstances on the roads, but also to achieve more efficient traffic management. Security  in VANETs faces many challenges due to the open broadcasting of wireless communications and the high-speed mobility of  vehicles. In these networks, any malicious misbehaving user that can inject  false information,  or modify/replay any previously  disseminated message, could be fatal to the others. 

VANETs are considered a promising research area of mobile communications because they offer a wide variety of possible applications, ranging from the aforementioned road safety and transport efficiency, to commercial services, passenger comfort, and infotainment delivery. Furthermore, VANETs can be seen as an extension of mobile ad-hoc networks where there are not only mobile nodes, named On-Board Units (OBUs), but also static nodes, named Road-Side Units (RSUs). The so-called Intelligent Transportation System (ITS) includes two  types of communications: 
\begin{itemize}
\item between OBUs: Vehicle TO Vehicle  (V2V)
\item between OBUs and RSUs: Vehicle TO Infrastructure (V2I)  and  Infrastructure TO Vehicle (I2V). 
\end{itemize}

It is crucial to have a good system for revocation of malicious vehicles to protect the safety of other users. Some of the most useful applications in VANETs require efficient revocation systems. One of these applications is the one  highlighted in \cite{fmf.surveyVANET}, based on information from other vehicles, which relies on V2V, V2I and I2V to run various applications and to exchange event-driven or periodic messages. Other highlighted applications are based on:

\begin{itemize}
\item \textit{Cooperative forward collision warning}. This system accomplishes the goals necessary to assist a vehicle in avoiding becoming involved in an accident with the vehicle travelling ahead of it. The system uses V2V communication with  multi hop relaying in order to send warning messages.
\item \textit{Vehicle-based road condition warning}.  Vehicles collect  information about  road conditions via their sensors, and after collecting sufficient information at their OBUs, they process these data to determine the road situation in order to send warning messages to other vehicles.
\item \textit{Cooperative collision warning}. The main goal of this application is to warn the driver about any predicted accident, by using V2V communication.
\end{itemize}

All these applications require an accurate and reliable system to exchange messages between vehicles. The receiver first must verify the reliability of the sender in order to know whether the received information is trustable or not. Therefore, an effective system of verification of revoked users is necessary.

Both the European standard for ITS, named ITS-G5, and its American counterpart, named Wireless Access in Vehicular Environment (WAVE), are based on the IEEE 802.11p amendment to the IEEE 802.11 standard. 
The WAVE standard on the so-called Dedicated Short-Range Communications (DSRC) channels is defined as the only wireless technology that can potentially meet the extremely short latency requirement for road safety messaging and control under any condition because DSRC was specifically designed for automotive use. According to DSRC/WAVE, vehicles periodically exchange  with nearby vehicles beacons containing  sender's  information such as location and speed because many VANET applications, such as  cooperative collision warning, rely on the information embedded in these beacons. 

Within the family of standards for vehicular communications IEEE 1609  based on the  IEEE 802.11p, the standard  1609.2 deals in particular with the issues related to security services for applications and management messages. This standard describes  the use of PKIs, CAs and CRLs, and implies that in order to revoke a vehicle, a CRL has to be issued by the CA to the RSUs, who are in charge of sending this information to the OBUs. 
Each vehicle is assumed to have a pair of keys: a private signing key and a public verification key certified by the CA; and any VANET message must contain: a timestamp with the creation time, the sender's signature, and the sender's public-key certificate. In particular,  the IEEE 1609.2 standard proposes both broadcast authentication and non-repudiation through the use of the elliptic curve digital signature algorithm. 

In order to protect  privacy in VANETs, each OBU can obtain multiple certified key pairs  and use different  public keys each time. These public keys are linked to  pseudonyms  that allow preventing location tracking by  eavesdroppers. Therefore, once VANETs are implemented in practice on a large scale, the size of CRLs will grow rapidly due  to the increasing number of OBUs and to  the use of such multiple   pseudonyms. Thus, it is foreseeable that if CRLs are used, they will become extremely large and unmanageable. Moreover,  this context can bring a phenomenon known as implosion request, consisting of several nodes who synchronously want to download the CRL at the time of its updating, producing serious congestion and overload of the network, which could ultimately lead to a longer latency in the process of validating a certificate.

The proposal described in this paper  proposes the use of  hash trees to achieve cooperative revocation of malicious users  both in certificate-based and in IBC-based authentication in VANETs. In particular, it uses  a $k$-ary hash tree as an ADS for revocation management. By using this ADS, the process of query on the validity of public certificates/pseudonyms will be more efficient because OBUs will send queries to RSUs, who will answer them on behalf of the TTP. In this way, at the same time this TTP will no longer be a bottleneck, and OBUs will not have to download any entire CRL. Instead of that, they will have to manage hash trees where the leaf nodes contain revoked certificates/pseudonyms. In particular, the used  $k$-ary trees are based on the application of a duplex construction of the Secure Hash Algorithm SHA-3 recently chosen as standard, because the combination of both structures allows improving efficiency of updating and querying  revoked certificates/pseudonyms.

This paper is organized as follows. Section 2 addresses the general problem of the use of certificate revocation lists in VANETs, and provides a succinct revision of related works. Then, Section 3 introduces  the necessary preliminaries and a brief explanation of the tree-based  proposal. Afterwards, Section 4 includes the specific descriptions of the algorithms to operate with the proposed tree, and focuses on a performance evaluation based on results of their implementations and simulations. Finally, Section 5 discusses conclusions and possible future research lines.

\section{Related Works}
\label{Related Works}

In many applications, the use of public-key cryptography is essential for information security \cite{2}. In those cases,  the revocation problem is one of the most difficult to solve.

Usual revocation procedures are  based on a CA that manages revoked public-key certificates by including  the corresponding certificate serial numbers in a CRL and distributing this CRL within the network in order to let users know which nodes are no longer trustworthy \cite{NN2000}. Under these circumstances, it is very important that the distribution of the CRL is done efficiently in order to allow  that the knowledge about untrustworthy nodes can be spread quickly to the entire network. 

As aforementioned, the family of standards IEEE 1609 describes the use of PKI in VANETs. In particular, the work \cite{9} defines a proposal for the use of a PKI to protect messages and mutually authenticate entities in VANETs. As a continuation of that work, the paper \cite{19} introduces a PKI-based security protocol where each vehicle preloads anonymous public/private keys and a TTP stores all the anonymous certificates of all the vehicles. This scheme cannot be considered efficient in the certificate management process.

Also based on a PKI, a well-known solution for strong authentication in VANETs is based on the signature of each message \cite{11}. However, the use of a traditional approach to PKIs may fail to satisfy the real time requirement in vehicular communications because according to the DSRC protocol, each OBU will periodically transmit beacons so even in a normal traffic scenario, it is a very rigorous requirement to deploy an authentication scheme that allows at the same time efficient revocation of invalid public keys, and efficient use of valid public keys, which is exactly the main goal of this work.

For revocation in VANETs, previous works assume that the entire CRL may be delivered by broadcasting it directly from RSUs  to  OBUs \cite{JD08}, and then distributed among OBUs  cooperatively \cite{Cooperation}. However, the large size of VANETs, and consequent large size of the CRLs, makes this approach infeasible due to the overhead it would cause to network communications. This issue is further increased with the use of multiple pseudonyms for the nodes, what has been suggested to protect privacy and anonymity of OBUs \cite{PB08}. 

Since there are almost one thousand million cars in the world \cite{M2000},  considering the use of pseudonyms, a direct conclusion is that the number of revoked certificates might reach soon the same amount, one thousand million. On the other hand, assuming that each certificate takes at least 224 bits, in such a case the CRL size would be 224 Gbits, what means that its management following the traditional approach would not be efficient. Even though regional CAs were used and the CRLs could be reduced to 1 Gbit, by using the 802.11a protocol to communicate with RSUs in range, the maximum download speed of OBUs would be between 6 and 54 Mbit/s depending on vehicle speed and road congestion, so on average an OBU would need more than 30 seconds to download a regional CRL from an RSU. 

A straight consequence of this size problem is that a new CRL cannot be issued very often, what would affect the freshness of revocation data. On the other hand, if a known technique for large data transfers were used for CRL distribution as solution for the size problem, it would result in higher latencies, what would also impact in the revocation data validity. Consequently, a solution not requiring the distribution of the full CRL from RSUs to OBUs, like the one proposed in this work, would be very helpful for the secure and efficient operation of VANETs.

A general revocation method not based on CRLs, called Online Certificate Status Protocol (OCSP) \cite{Myers}, involves a multitude of validation agents that respond to client queries with signed replies indicating the current status of a target certificate. This explicit revocation method has an unpleasant side effect because it divulges too much information. Since validation agents constitute a global service, they must involve enough replication to handle the load of all validation queries, what means that the signature key must be replicated across many servers, which is either insecure or expensive. 

Another general solution not based on CRLs, called Certificate Revocation Tree (CRT), was proposed in \cite{15} as an improvement of OCSP involving a single highly secure entity that periodically posts a signed CRL like data structure to many insecure validation agents so that users query these agents. In CRTs, the leaf nodes are statements concerning revoked certificates, and the CA signs the root. By using CRTs, the responder can prove the status of any certificate by showing the path from the root to the leaf node without signing the response, because the signatures of any leaf node are identical, and given by the signature contained in the root. Thus, no trust in the responder is necessary. The proposal here described is based on this idea.

In general, a hash tree is a tree structure whose nodes contain digests that can be used to verify larger pieces of data \cite{M1982}. The leaf nodes in a hash tree are hashes of data blocks while nodes further up in the tree are the hashes of their respective children so that the root of the tree is the digest representing the whole structure.  Hash trees usually require the use of a cryptographic hash function in order to prevent collisions. Most  implementations of hash trees are binary, but this work proposes the use of the more general structure of k-ary trees because when combining it with a particular choice of cryptographic hash function, it is possible to optimize the update of the hash tree. 

The basic ADS proposed in \cite{15} is a Merkle hash tree \cite{hashtree} where the leaf nodes represent revoked certificates sorted by serial number. A client sends a query to the nearest agent, which produces a short proof that the target certificate is (or not) on the CRT. The work\cite{12} introduces several methods to traverse Merkle trees allowing time space trade-offs. Other ADSs based on multi-dimensional tree structures are studied in \cite{17} to support efficient search queries, allowing the retrieval of authenticated certificates from an untrusted repository used for dissemination by various credential issuers. Besides, many tree-balancing algorithms have been proposed in the bibliography for hash trees \cite{4}. For instance, AVL trees are balanced by applying rotation, B-trees are balanced by manipulating the degrees of the nodes, and 2-3 trees contain only nodes with at least 2 and at most 3 children. However, in the particular application of public-key revocation, balancing trees does not necessary minimize the overall communication. 

Another interesting problem with CRTs appears each time a certificate is revoked as the whole tree must be recomputed and restructured. Skip-lists proposed in \cite{7} \cite{8} can be seen as a natural and efficient structure to reduce communication by balancing the CRT. However, they are not good solutions for other problems such as insertion of new leaf nodes. 

Hash trees are usually based on widely used hash functions. This paper proposes the use of a new version of Keccak as  cryptographic hash function in the hash tree. Keccak is the cryptographic hash function used in the new  SHA-3 standard \cite{SHA3}. The requirements set by NIST for SHA-3 candidates included typical security properties of hash functions, such as collision resistance, preimage resistance and second preimage resistance \cite{Andreeva}. Different types of implementations of SHA-3 finalists have been evaluated in several works \cite{Guo} \cite{Aoki}, obtaining in most cases positive conclusions. The original SHA-3 uses a sponge construction \cite{Keccak}, which in a cryptographic context is an operating mode on the base of a fixed length transformation and a padding rule. Instead of it, this paper proposes a duplex construction \cite{duplexConstruction}. The main advantage of the duplex construction is that  it provides digests on the input blocks received so far, what is applied in the proposal here described so that the hash tree is constructed efficiently. 

Previous proposals that can be considered close to this work are \cite{fmf.mht} \cite{fmf.coach} because they use hash trees for revocation in VANETs. However, they use neither perfect k-ary trees nor a specific hash to optimize tree operations.

\section{Tree-Based Proposal}
\label{Revocation Tree}

The scheme described in this paper is based on  using as ADS a k-ary tree, which  is a rooted tree where each node has no more than $k$ children. The use of k-ary hash trees instead of binary trees allows increasing efficiency of the construction and update of hash trees. Thus, one of the major drawbacks of ordered tree structures, which is the necessary reconstruction when there are changes in the tree, only occurs when the k-ary tree requires a new level of depth, because otherwise the nodes simply are inserted from left to right to complete the level of depth corresponding to their query frequency. In this way, our proposal is based on a dynamic tree-based data structure that varies depending on the number of revocations.

The proposed model  is based on the following notation:

\begin{itemize}
\item $h$: Cryptographic hash function used in the hash tree.
\item $D$ ($\geq 1$):  maximum Depth of the hash tree.
\item $d$ ($< D$): Depth of an internal node in the hash tree.
\item $s$: Number of revoked certificates/pseudonyms.
\item $R_j$ ($j=1, 2,..., s$): Serial number of the $j-th$ Revoked certificate/pseudonym.
\item $N_{ij}$ ($i=D-d$ and $j=0, 1...$): Internal Node of the hash tree obtained by hashing the concatenation of all the digests contained in its children.
\item $N_{0j}$ ($j=0, 1...$): Leaf node of the hash tree containing $h(R_j)$, ordered  according to  revocation.
\item $k$: Maximum number of children for each internal node in the hash tree.
\item $f$: Keccak function used in SHA-3.
\item $n$: Bit size of the digest of $h$, which is here assumed to be the lowest possible size of SHA-3 digest, 224.
\item $b$: Bit size of the input to $f$, which is here assumed to be one of the possible values of Keccak, 800.
\item $r$: Bit size of input blocks after padding for  $h$, which is here assumed to be 352.
\item $c$: Difference between $b$ and $r$, which is here assumed to be as in SHA-3, $2n$, that is 448.
\item $l$: Bit size of output blocks for building the digest of $h$, which is here assumed to be lower than $r$.
\end{itemize}

\begin{figure}
	\centering
		\includegraphics[width=3.5in]{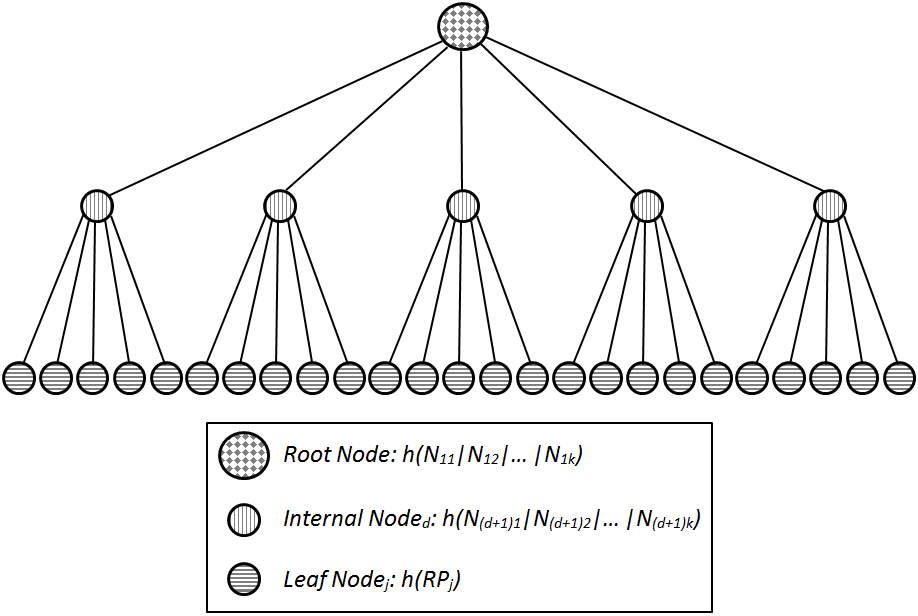}
	\caption{Hash Tree Based on a 5-ary Tree}
	\label{fig.fmf.5arytree}
\end{figure}

In order to build the tree, the first parameter to consider is the maximum number of children per node. This parameter defines the $k$ of the $k$-ary tree to be built. If $k$  equals 2, the resulting tree is the typical binary tree, but the proposal allows different values for $k$, such as 3, 4, 5, etc. For instance, a 5-ary tree is shown in Figure \ref{fig.fmf.5arytree}.

In order to learn how to find a node in the tree, a hash table is used to map each revoked certificate/pseudonym with the exact path that defines the tree.

In the proposed scheme, the bandwidth cost of sending the new versions of the revocation tree from the TTP to the OBUs is significantly reduced because only nodes in the tree that have changed need to be updated. This implies a significant improvement with respect to previous tree-based schemes for revocation management because one of their main problems is the necessary update of the entire hash tree every time a new leaf node is added or an existing leaf node is deleted.

The authenticity of the used hash tree structure  is guaranteed thanks to the TTP signature of the root. The procedure to follow when the part of the revocation tree that is necessary for authenticity verification is pushed from an RSU to an OBU after this latter queries the first one about a certificate/pseudonym, is as follows.  If the RSU finds the digest of the queried  certificate/pseudonym among the leaf nodes of the tree because it is  revoked, then the RSU sends to the OBU the route from the root to the corresponding leaf node, along with all the siblings of the nodes on this path. After checking all the digests corresponding to the received path and the TTP signature of the root, the OBU gets convinced of the validity of the evidence on the revoked certificate/pseudonym received from the RSU. 

\begin{figure}
	\centering
		\includegraphics[width=3.5in]{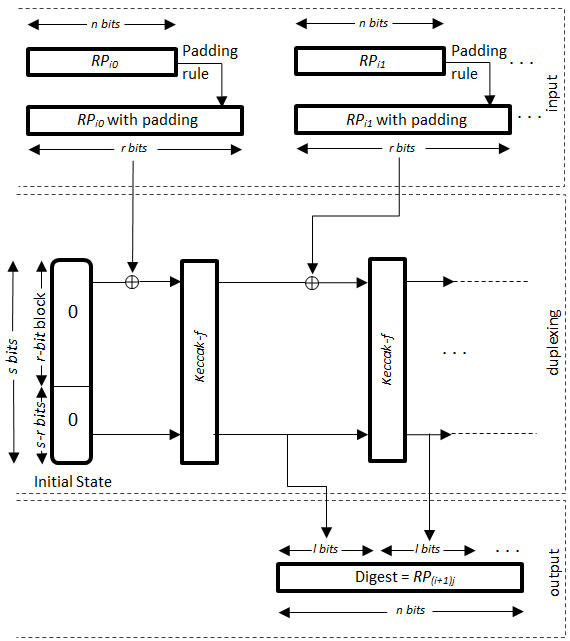}
	\caption{Proposed Duplex Construction}
	\label{fig:duplexConstruction}
\end{figure}

Regarding the cryptographic hash function $h$ used in the hash tree, this proposal is based on the use of a new version of the Secure Hash Algorithm SHA-3. In SHA-3,  the basic cryptographic hash function $f$ called Keccak contains 24 rounds of a basic transformation and its input is represented by a $5 \times 5$ matrix of 64-bit lanes. In contrast,  our proposal is based on 32-bit lanes. Another proposed variation of SHA-3 is the use of a duplex version of the sponge structure of SHA-3. On the one hand, like the sponge construction of SHA-3, the proposal based on a duplex construction also uses Keccak as fixed-length transformation $f$, the same padding rule and data bit rate $r$. On the other hand, unlike a sponge function, the duplex construction output corresponding to an input string might be obtained through the concatenation of the outputs resulting from successive input blocks (see Figure  \ref{fig:duplexConstruction}). 

The use of the duplex construction in the proposed hash tree allows the insertion of a new revoked certificate/pseudonym as new leaf of the tree by running a new iteration of the duplex construction only on the new revoked certificate/pseudonym. In particular, the RSU can take advantage of all the digests corresponding to the sibling nodes of the new node, which were computed in previous iterations, by simply discarding the same minimum number of the last bits of each one of those digests so that the total size of the resulting digest of all the children remains the same, $n$. This proposed procedure makes the hash tree construction more efficient than previous tree-based schemes when a new leaf corresponding to a new revoked certificate/pseudonym has to be inserted in the tree. While the maximum number of children of an internal node has not been reached, the RSU has to store not only all the digests of the tree structure but also the state resulting from the application of Keccak hash function $f$ in the last iteration corresponding to such internal node, in order to use it as input in a next iteration. 

Periodic deletions of certificates/pseudonyms that are in the tree and reach their expiration date, require  reconstruction of the part of the tree involving the path from those nodes to the root. Thus, in order to maximize the proposal, such tree  reconstruction is linked to the moment when all the sibling nodes of some internal node  expire because this  avoids  unnecessary reductions of the system efficiency by having to reconstruct the tree very often.

The TTP is  responsible for periodically updating the tree by deleting the expired certificates/pseudonyms, and for reconstructing the tree when necessary. After each update, the TTP sends the corresponding modifications of the updated tree to all RSUs.
The RSU has to search vehicle certificates/pseudonyms in the revocation tree each time an OBU requests it. The RSU must provide the requesting OBU either with a verifiable revocation proof of any revoked certificate/pseudonym or with a signed message indicating that the queried certificate/pseudonym has not been revoked and is labelled as 'OK'. In the first case, by using the answer data, the OBU can verify the TTP signature of the received signed root, recompute the root of the revocation tree, and check it by comparing it with the received signed root. 

When using ID-based cryptography, the  proposal  is based on the use of a group of pseudonyms set for each OBU, so that for each pseudonym the TTP provides the OBU with a corresponding private key. If any of these pseudonyms is revoked by the TTP, it inserts all the pseudonyms corresponding to the same OBU in the revocation tree.

\section{Performance Evaluation}
\label{Performance Evaluation}   

The proposal can be considered computationally efficient  because it obviates the need to sign each RSU's reply. In general, the proposal  does not require trusting all RSUs. Indeed, the only case when trust on RSU is necessary is when it provides an 'OK' answer because this could be a fraud. 

In this regard, when an OBU receives an 'OK' message signed by a cheating RSU, it trusts it momentarily. However, when it contacts another RSU, it asks it again about the same certificate/pseudonym. If this RSU provides the OBU with a proof of revocation whose timestamp contradicts the 'OK' answer signed by the questioned RSU, the OBU sends to the latter RSU an impeachment on the questioned RSU, so that the honest RSU can send it to the TTP, who will revoke its public key by deleting it directly from the revoked RSU. Otherwise, if the second RSU also sends a signed 'OK' message, the OBU goes on asking about the same certificate/pseudonym until it reaches either a contradiction or a prefixed trust threshold. 

Thus, each OBU stores locally in two separate and complementary structures, the pseudonyms of those OBUs that it has previously checked as unreliable, and of those OBUs that have been reliable till then. Therefore, in the future, if it reconnects with any of these vehicles, it can use such information to decide how to proceed. If there is no RSU nearby, it uses these data to decide whether to establish the communication or not. Otherwise, even if there is an RSU nearby, there is no need to re-ask it about a checked revoked certificate/pseudonym.

The choice of adequate values for the different parameters in our proposal must be done carefully, taking into account the relationships among them. In particular, since the maximum tree size:

\begin{center} 
$n(1+k+k^2+k^3+ \cdot \cdot \cdot + k^D)= \frac{n(k^{(D+1)}-1)}{k-1}$
\end{center} 

\noindent is upperbounded by the size of available memory in the RSU, and the maximum number of leaf nodes of the $k$-ary tree $k^D$ is lowerbounded by the number of revoked certificates/pseudonyms $s$,  both conditions can be used to deduce the optimal value for $k$.

The two following subsections contain respectively results obtained from different implementations and simulations of the proposal, and brief algorithmic descriptions of the tree operations. 

\subsection{Simulations}

In order to achieve a realistic evaluation, real data of vehicular environments and detailed studies have been used. Depending on the number of nodes in vehicular environments, the number of revoked certificates has been estimated using the proposal described in \cite{fmf.berkovits}. In particular, this statistical research of NIST estimated that 10\% of the certificates need to be revoked.

The data used in the comparisons have been chosen  according to the study presented in \cite{fmf.mobilityMadrid}. Such a research focuses on Madrid city and estimated its  fleet in  1.7 million vehicles. Using these data and the NIST study about revocation rate in VANETs.

A real scenario was simulated with the characteristics shown in Table \ref{table.fmf.simulationScenario}.

\begin{table}[h!]
  \centering
   \caption{Parameter Values for the Simulation Scenario}
	\begin{tabular}{ l  l }
		\hline
		\textbf{Parameter} & \textbf{Value} \\ \hline
		    Scenario & Madrid City (Spain, 2014) \\
			Size scenario & $25 Km^{2}$ \\
			Simulation time & 1000 s \\
			Number of vehicles & 1349 \\
			\% of Vehicles with OBU & {10, 20, 30, 40, 50, 60 , 70, 80, 90, 100} \\
			MAC & IEEE 802.11p \\
			Propagation model & DSDV \\
			Transport protocol & UDP \\
			Size package & 1 Kb \\
			Link Layer & LL \\
			\hline
	\end{tabular}
	\label{table.fmf.simulationScenario}
\end{table}

Features of vehicles in the simulations are defined in Table \ref{table.fmf.simulationOBU}.

\begin{table}[h!]
  \centering
   \caption{Vehicle and OBU Profile}
	\begin{tabular}{ l  l }
		\hline
		\textbf{Parameter} & \textbf{Value} \\ \hline
		    Speed & [0-33] m/s \\
			Transmission distance & 55 m \\
			Antenna & OmniAntenna \\
			txPower & 1.4 mW \\
			rxPower & 0.9 mW \\
			sensePower & 0.00000175 mW \\
			idlePower & 0 mW \\
			Initial Energy & 75 J \\
			\hline
	\end{tabular}
	\label{table.fmf.simulationOBU}
\end{table}

The simulations were done using multiple software packages in order to add credibility and to present a realistic simulation. Therefore, the scenario that has been used for simulations comprises a real traffic situation in the city of Madrid (Spain) in 2014. 

For the traffic generation, SUMO software tool \cite{fmf.sumo} has been used. SUMO is a free and open traffic simulation suite that allows modelling of intermodal traffic systems, including road vehicles, public transport and pedestrians. 

In order to simulate the architecture and communications of a VANET,  a tool called NS-2 \cite{fmf.ns2}  was used. NS-2 is a discrete event simulator targeted at networking research that provides substantial support for simulation of TCP, routing, and multicast protocols over wired and wireless (local and satellite) networks.

The  interaction between the traffic generated with SUMO and the  network simulated with NS-2 is generated  using MOVE, which allows users to rapidly generate realistic mobility models for VANET simulations. MOVE is built on top of an open source micro-traffic simulator SUMO. Its output  is a realistic mobility model that can be immediately used with popular network simulators, such as NS-2. 

\begin{figure}[h!]
	\centering
		\includegraphics[width=3.5in]{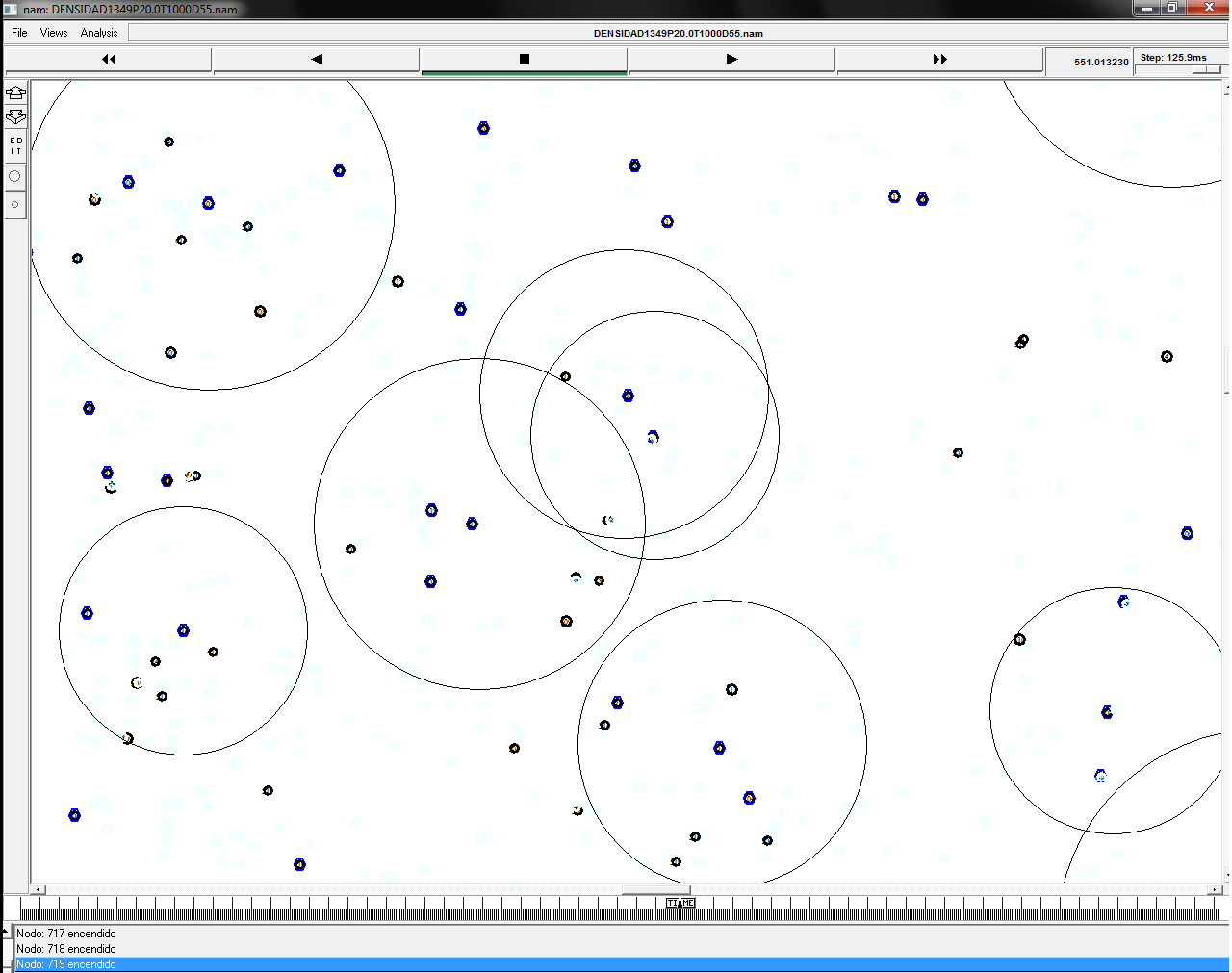}
	\caption{Architecture and Communications of  VANET Simulation in NS-2}
	\label{fig.fmf.ns2}
\end{figure}

As aforementioned, NS-2 was used for the simulation of the network. The data generated with NS-2 were subsequently transferred to SUMO for their display on the map.  Sent messages and interactions between nodes in NS-2 can be seen in Figure \ref{fig.fmf.ns2}.

\begin{figure}[h!]
	\centering
		\includegraphics[width=3.5in]{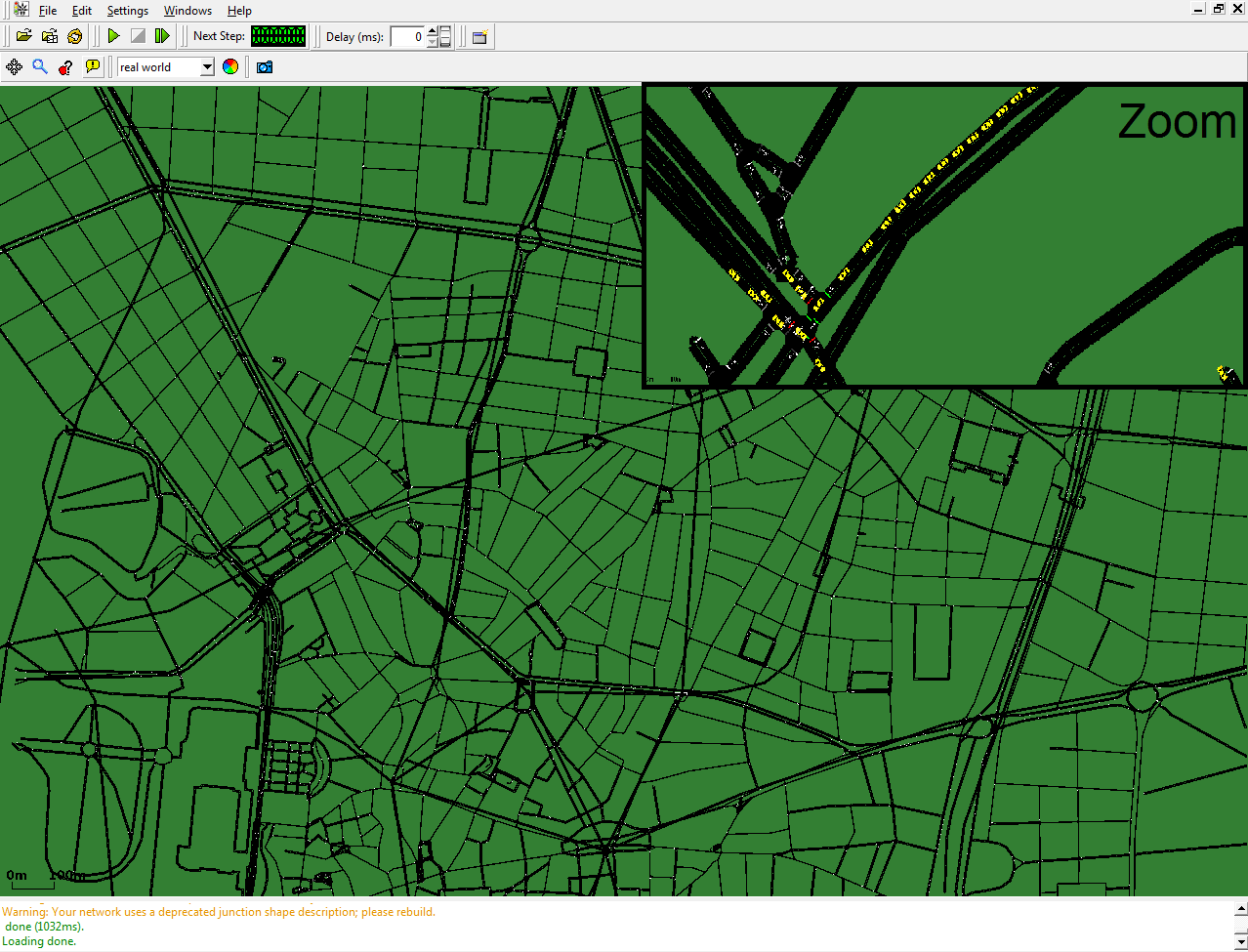}
	\caption{Example of SUMO Traffic Simulation}
	\label{fig.fmf.sumo}
\end{figure}

With the obtained simulation data and settings shown in previous tables, the visual aspect of SUMO software corresponding to the selected scenario is shown in Figure \ref{fig.fmf.sumo}.

\begin{figure}[h!]
	\centering
		\includegraphics[width=3.5in]{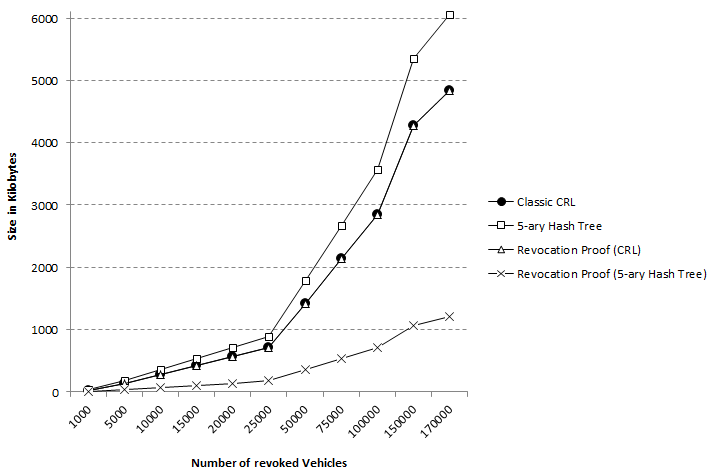}
	\caption{Comparison Between  Typical Revocation Lists and Our Proposal}
	\label{fig.fmf.sizeComparison}
\end{figure}

From the comparison between  the size of the typical revocation list and the size of the proposed revocation tree, the obtained results can be seen in Figure \ref{fig.fmf.sizeComparison}. It shows that although the proposed revocation structure uses more space than the usual revocation list, the revocation proof requires much less space. This is because when using typical revocation lists each vehicle obtains the full list each time it requires it, while in our scheme it only gets a path in the hash tree when it is necessary. As RSUs are assumed to  have sufficient memory, their need to store the complete hash tree is not a problem. The key issue to optimize is the shipping of the revocation proof because  vehicles move at high speeds and typical revocation lists  are too heavy to be sent in these environments.

\begin{figure}[h!]
	\centering
		\includegraphics[width=3.5in]{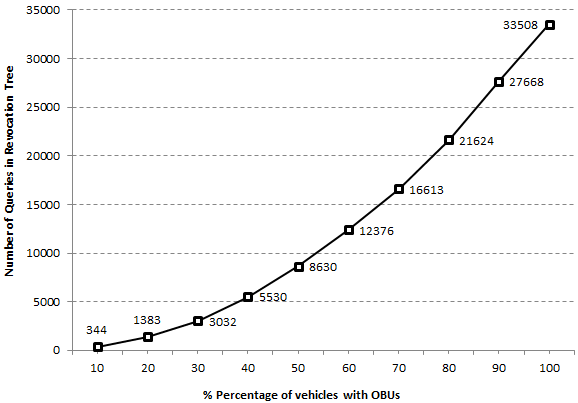}
	\caption{Queries in the Simulated Scenario}
	\label{fig.fmf.queries}
\end{figure}

The number of authentications and  queries, together with the  revocation tree designed and implemented in this paper are represented in Figure \ref{fig.fmf.queries}, depending on the percentage of vehicles with OBUs in the described scenario.

\begin{figure}[h!]
	\centering
		\includegraphics[width=3.5in]{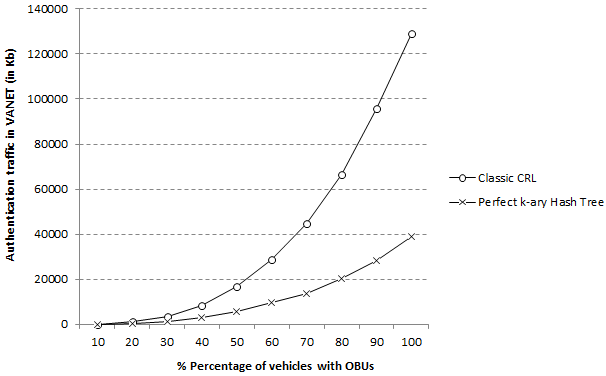}
	\caption{Traffic Generated in the Verification Process of Revoked Certificates/Pseudonyms}
	\label{fig.fmf.sizekary}
\end{figure}

Traffic connections and authentications  generated in the VANET  have been estimated at random in order to check node revocation. This estimation was drawn up on the simulation scenario and compared with the traffic generated by conventional revocation lists on the same scenario. The comparison results, which demonstrate the efficiency of our proposal can be seen in Figure \ref{fig.fmf.sizekary}.

\subsection{Tree Algorithms}

To optimize operations in the tree, have developed specific algorithms that work on perfect k-ary hash trees. It has been designed and implemented an algorithm to find a revoked pseudonym in the tree (see Algorithm \ref{alg:search}). It has designed a specific algorithm to insert a new pseudonym revoked in the tree. So that only a single iteration is required in the hash proposed parent node to recalculate the function, where the new node is inserted (see Algorithm \ref{alg:insert}). An algorithm for removing a revoked pseudonym tree is proposed (see Algorithm \ref{alg:delete}). Only leaf nodes of the tree will be deleted when they expire. Finally, a fast and efficient algorithm for the restructuring of the tree is proposed (see Algorithm \ref{alg:reconstruct}). A tree is to be restructured when the deletion or insertion of a node causes the removal or creation of a depth level.

\vspace{0.3cm}

\begin{algorithm}[h!]
 \caption{Search Algorithm} \label{alg:search}
 \KwIn{\textit{rpSearch}, Tree-ID of leaf node to be searched.}
 \KwOut{\textit{retPath}, Path from root to leaf node.}
 Add Root Node  to \textit{retPath}\;
 \For{$i \gets (D-1)$ \textbf{to} $0$} {
    Compute \textit{rpSearch} branch to $i$ height\;
    Add Siblings Nodes of the computed top level branch  to \textit{retPath}\;
 }
 Compute \textit{rpSearch} Leaf Node\;
 Add \textit{rpSearch} Leaf Node and Sibling Nodes to \textit{retPath}\;
 return \textit{retPath}\;
\end{algorithm}

\vspace{0.3cm}

\begin{algorithm}[h!]
 \caption{Insertion Algorithm} \label{alg:insert}
 \KwIn{\textit{rpNew}, Tree-ID of Leaf Node to be inserted.}
 Go to Last Depth Level where  the \textit{rpNew} Node must be inserted\;
 \eIf{It is necessary to Create a New Depth Level}{
 	Generate New Depth Level\;
 	Insert \textit{rpNew} Node\;
 	Reconstruct Hash Tree\;
 } {
    Insert \textit{rpNew} Node\;
 }
\end{algorithm}

\vspace{0.3cm}

\begin{algorithm}[h!]
 \caption{Deletion Algorithm} \label{alg:delete}
 \KwIn{\textit{rpDelete}, Tree-ID of Leaf Node to be deleted.}
 Compute \textit{rpDelete} Parent Node\;
 Delete \textit{rpDelete} Node\;
 \If{\textit{rpDelete} Parent Node has no leaf node}{
 	Delete \textit{rpDelete} Parent Node\;
 }
 Reconstruct Hash Tree\;
\end{algorithm}

\vspace{0.3cm}

\begin{algorithm}[h!]
 \caption{Reconstruction Algorithm} \label{alg:reconstruct}
 \KwIn{\textit{Tree}, Input Hash Tree.}
 \KwOut{\textit{Tree}, Reconstructed Hash Tree.}
 \For{$i \gets 1$ \textbf{to} $D$} {
 	Compute the branch of \textit{i}\ on the \textit{Tree}\;
 	Relocate the nodes of the branch\ on the \textit{Tree}\;
 }
 \For{$i \gets (D-1)$ \textbf{to} $0$} {
    \For{$j \gets 0$ \textbf{to} $Number\_of\_Nodes\_at\_Level\_i$} {
    	Recompute Hash Value of the Node \textit{j} from its Children at Level \textit{i+1}\;
 	}
 }
 return \textit{Tree}\;
\end{algorithm}

\section{Conclusions}

Revocation of malicious users in vehicular ad-hoc networks is a complex and critical problem. This work proposes a new way to manage revocation in these networks  when using certificate-based  or identity-based authentication. In particular, to improve the performance of revocation lists, this paper proposes the use of a data structure based on authenticated dynamic hash $k$-ary trees. $k$-ary hash trees are used here in conjunction with a new  duplex version of the standard SHA-3 hash function to improve performance of revocation through  algorithms optimized to simplify both the search of queried certificates/pseudonyms and the update of the trees when inserting new revoked certificates/pseudonyms. Consequently, the proposal can be considered especially useful both in urban environments and when vehicular ad-hoc networks are widely deployed because then revocation lists will be frequently updated. Although this is a work in progress, the results of the simulations are very promising.

\section*{Acknowledgment}
Research supported by the Spanish MINECO and the European FEDER under projects TIN2011-25452, BES-2012-051817 and IPT-2012-0585-370000, and the FPI scholarship BES-2012-051817.


\begin{thebibliography}{22}

\bibitem{fmf.surveyVANET}
S.~Al-Sultan, M.M.~Al-Doori, A.H.~Al-Bayatti, H.~Zedan,  
A comprehensive survey on vehicular Ad Hoc network,
Journal of Network and Computer Applications, pp. 380-392, 2014.

\bibitem{2}	S. Blake-Wilson, Information security, mathematics, and public-key cryptography, Designs, Codes and Cryptography 19(2-3), pp. 77-99, 2000.

\bibitem{NN2000} M.~Naor, K.~Nissim, Certificate revocation and certificate update, IEEE Journal on Selected Areas in Communications 18(4), pp. 561-570, 2000.

\bibitem{9}	J.P. Hubaux, S. Capkun, J. Luo, The security and privacy of smart vehicles, IEEE Security and Privacy 2(3), pp. 49-55, 2004.

\bibitem{19}	M. Raya, J.P. Hubaux, Securing vehicular ad hoc networks,  Computer Security 15(1), pp. 39-68, 2007

\bibitem{11}	IEEE-1609, Family of standards for Wireless Access in Vehicular Environments (WAVE), US Department of Transportation, 2006.

\bibitem{JD08} D.~Jiang, L.~Delgrossi, IEEE 802.11 p: Towards an international standard for wireless access in vehicular environments,  IEEE Vehicular Technology Conference VTC Spring, pp. 2036-2040, 2008.

\bibitem{Cooperation} J.~Molina-Gil, P.~Caballero-Gil, C.~Caballero-Gil, Enhancing Cooperation in Wireless Vehicular Networks, International Workshop on 
Security in Information Systems, pp. 91-102, 2011.

\bibitem{PB08} P.~Papadimitratos, L.~Buttyan, T.~Holczer, E.~Schoch, J.~Freudiger, M.~Raya, Z.~Ma, F.~Kargl, A.~Kung,  J.-P.~Hubaux, Secure vehicular communications: Design and architecture, IEEE Communications Magazine 46(11), pp. 2–8,  2008.

\bibitem{M2000} A.J.~McMichael, The urban environment and health in a world of increasing globalization: issues for developing countries, Bulletin of the World Health Organization 78(9), pp. 1117-1126, 2000.

\bibitem{Myers}
Myers, M., Ankney, R., Malpani, A., Galperin, S.,  Adams, C. (1999). X. 509 Internet public key infrastructure online certificate status protocol-OCSP. RFC 2560.

\bibitem{15}	P. Kocher, On certificate revocation and validation,  Financial Cryptography, Lecture Notes in Computer Science 1465, pp. 172-177, 1998.

\bibitem{M1982} R.C.~Merkle, Method of providing digital signatures,  U.S. Patent No. 4,309,569, 1982.

\bibitem{hashtree} R.C.~Merkle, Protocols for public key cryptosystems, IEEE Symposium on Security and privacy 1109, pp. 122-134, 1980.

\bibitem{12}	M. Jakobsson, T. Leighton, S. Micali, M. Szydlo, Fractal merkle tree representation and traversal, CT-RSA, Lecture Notes in Computer Science 2612, pp.314-326, 2003.

\bibitem{17}	V. Miller, Short programs for functions on curves,  Unpublished manuscript, 97, pp. 101-102, 1986.

\bibitem{4}	T. Cormen, C. Leiserson, R. Rivest, Introduction to algorithms, MIT Press, 1990.

\bibitem{7}	M. Goodrich, M. Shin, R. Tamassia, W. Winsborough, Authenticated dictionaries for fresh attribute credentials, Trust Management, Lecture Notes in Computer Science 2692, pp. 332-347, 2003.

\bibitem{8} 	M. Goodrich, R. Tamassia, N. Triandopoulos, R. Cohen, Authenticated data structures for graph and geometric searching, CT-RSA, Lecture Notes in Computer Science 2612, pp. 295-313, 2003. 

\bibitem{SHA3} V.~Rijmen, Extracts from the SHA-3 Competition, Selected Areas in Cryptography, pp. 81-85, 2013.

\bibitem{Andreeva} E. Andreeva, B. Mennink, B. Preneel, M. Skrobot, Security analysis and comparison of the SHA-3 finalists BLAKE, Grostl, JH, Keccak, and Skein,  AFRICACRYPT, pp. 287-305, 2012.

\bibitem{Guo} X. Guo, M. Srivastav, S. Huang,  D. Ganta, M.B. Henry, L. Nazhandali, P. Schaumont, ASIC implementations of five SHA-3 finalists,  IEEE Design, Automation and Test in Europe Conference and Exhibition, pp. 1006-1011, 2012.

\bibitem{Aoki} K. Aoki,  K. Matusiewicz, G. Roland, Y. Sasaki, M. Schlaffer,  Byte Slicing Grostl: Improved Intel AES-NI and Vector-Permute Implementations of the SHA-3 Finalist Grostl, International Conference on E-Business and Telecommunications, pp. 281-295, 2012.

\bibitem{Keccak} G.~Bertoni, J.~Daemen, M.~Peeters, G.~Van Assche, Keccak sponge function family main document version 2.1, Updated submission to NIST (Round 2), 2010.

\bibitem{duplexConstruction} G.~Bertoni, J.~Daemen, M.~Peeters, G.~Van Assche, Duplexing the Sponge: Single-Pass Authenticated Encryption and Other Applications, Selected Areas in Cryptography, pp. 320-337, 2011.

\bibitem{fmf.mht}
J.~Munoz, O.~Esparza, C.~Ganan, J.~Mata-Diaz, J.~Alins, I.~Ganchev,  
MHT-based mechanism for certificate revocation in VANETs,
Wireless networking for moving objects: protocols, architectures, tools, services and applications, Springer, pp. 282-300, 2014.

\bibitem{fmf.coach}
C.~Ganan, J.~Munoz, O.~Esparza, J.~Mata-Diaz, J.~Alins,  
COACH: COllaborative certificate stAtus CHecking mechanism for VANETs,
Journal Network Computer Applications 36 (5), pp. 1337-1351, 2013.

\bibitem{fmf.berkovits}
S.~Berkovits, S.~Chokhani, J.~Furlong, J.~Geiter, J.~Guild,  
Public key infrastructure study: final report,
Technical Report, MITRE Corporation for NIST, 1995.


\bibitem{fmf.mobilityMadrid}
D. G. de Traficco. Portal estadistico. vehiculos. parque. http://www.dgt.es/, Accessed: 2015. 

\bibitem{fmf.sumo}
SUMO,  
http://www.dlr.de/ts/en/desktopdefault.aspx/tabid-9883/16931\_read-41000/,
[accessed: 25/03/2015].

\bibitem{fmf.ns2}
ns-2,  
http://www.isi.edu/nsnam/ns/,
[accessed: 25/03/2015].

\end{thebibliography}
\end{document}